\documentclass[usenatbib,useAMS]{mn2e}
\usepackage{epsfig}

\def\msun{{\rm M_{\odot}}}

\title [Maximum cluster densities ]
{Accretion-driven core collapse and
the collisional formation of massive stars}
\author[C.J.~Clarke \& I.A. ~Bonnell]{C.J.~Clarke$^1$, I.A. Bonnell, $^2$ \\
$^1$Institute of Astronomy, Madingley Rd, Cambridge, CB3 0HA, UK \\
$^2$ School of Physics \& Astronomy, University of St. Andrews, North Haugh, St. Andrews, Fife KY16 9SS, UK} 
\date{Submitted: July 2103}

\pagerange{\pageref{firstpage}--\pageref{lastpage}} \pubyear{2003}

\begin{document}
\def\lta{\mathrel{\spose{\lower 3pt\hbox{$\mathchar"218$}}
     \raise 2.0pt\hbox{$\mathchar"13C$}}}
\def\gta{\mathrel{\spose{\lower 3pt\hbox{$\mathchar"218$}}
     \raise 2.0pt\hbox{$\mathchar"13E$}}}
\def\Msun{{\rm M}_\odot}
\def\msun{{\rm M}_\odot}
\def\Rsun{{\rm R}_\odot}
\def\Lsun{{\rm L}_\odot}
\def\19{GRS~1915+105}
\label{firstpage}
\maketitle

\begin{abstract}
We consider the conditions required for a cluster core to shrink, by
adiabatic accretion of gas from the surrounding cluster, to densities
such that stellar collisions are a likely outcome. We show that 
the maximum densities attained, and hence the viability of collisions,
depends on a competition between core shrinkage (driven by accretion)
and core puffing up (driven by relaxation effects). The expected number of
collisions scales as $N_{core}^{5/3} \tilde v^2$ where $N_{core}$ is the
number of stars in the cluster core and $\tilde v$ is the free fall velocity
of the parent cluster (gas reservoir). Thus whereas collisions are very
unlikely in a relatively low mass, low internal velocity system such as the
Orion Nebula Cluster, they become considerably more important at the mass
and velocity scale characteristic of globular clusters. Thus stellar
collisions in response to accretion induced core shrinkage remains a
viable prospect in more massive clusters, and may contribute to the
production of intermediate mass black holes in these systems.
\end{abstract}


\section{Introduction}

  There has been considerable speculation in recent years that the
cores of clusters may be particularly conducive to the formation
of very massive stars, and their remnants. It is well known that
the most massive stars in young clusters tend to be concentrated
towards the centre (e.g. Hillenbrand \& Hartmann 1998, Sirianni et al
2002, Chen et al 2007)
and that, at least in some cases, this mass
segregation must be primordial (i.e. not purely the effect of 
two body relaxation in a cluster with an initially  uniform stellar
mass spectrum: see Bonnell \& Davies 1998, McMillan et al 2007 for
investigations of this issue). There is also some  observational
evidence for massive ($< 10^3-10^4 M_\odot$) remnants in some
old globular clusters (Gerssen et al 2003, van der Marel 2004, Noyola et al 2006), 
which would be consistent with these
objects following the same relationship between black hole mass
and host system mass as is found in galactic bulges.  Such intermediate
mass black holes in globular clusters could either represent the evolutionary
end-products of very massive stars in the cluster core or else could
have been formed through the merger of a number of black holes of
more modest mass (i.e. in the normal stellar regime), themselves being
the end-products of stellar evolution of massive stars in the cluster core
(van der Marel 2004, G\"urkan et al 2004).
In either case, it is important to know what is the mass spectrum of massive
stars in the cluster core -- whether it can be described purely by a
normal (Salpeter) IMF modified by dynamical mass segregation effects, or 
whether there are other processes that enhance the production of massive
stars in the cores of dense star clusters.

 One possibility is that the cores of clusters pass through a phase
of such exceedingly high density ($\sim 10^8$  pc$^{-3}$), that there is
an episode of runaway stellar collisions which may -
provided,  that the velocity 
dispersion in the cluster core remains low enough ($< 500$ km s$^{-1}$)  -
lead to the production of a  very massive star. The densities required for
stellar collisions far exceed the highest stellar densities ever observed,
suggesting that any putative ultra-dense phase is short-lived; the
associated mass density also far exceeds that of the  densest observed
gas in star forming clouds, implying that such a high density could {\it
not} be primordial. It is thus necessary to postulate a dynamical
mechanism that can drive the cores of young star clusters to the
requisite high densities.

  Two mechanisms have been proposed for this. The first is a purely
stellar dynamical  effect associated with a mass segregation 
instability (Spitzer 1969)in a cluster containing a range of stellar masses
(Rasio et al 2003, G\"urkan et al 2004). The
second is associated with an earlier evolutionary stage, when the
cluster is gas rich, and relies on the adiabatic shrinkage of a cluster
core that is accreting gas inflowing from larger radius in the cluster
( Bonnell, Bate and Zinnecker 1998). 

 In this Letter, we consider the issue
of what limits the stellar densities attainable in this latter case, 
and how this
maximum density relates to the parameters of the parent cluster. We will
present simple analytic arguments (which are consistent with the results of
existing numerical calculations) in order to show that the central density
may be driven to densities that exceed the mean cluster density by a factor
that scales with the square of the mass of the parent cluster.
We then go on to demonstrate that this places constraints on the parameters
of clusters in which stellar collisions may become important. 
We find that 
collisions are favoured in more massive systems since these shrink further,
in response to accretion, before this is reversed by two body effects.

\section{The evolution of a cluster core subject to accretion}

\subsection{Preamble}

   We develop some simple analytical arguments in which the system
is approximated by  a parent gas reservoir (`the cluster')
and a dense stellar core whose density increases with time due to adiabatic
accretion of gas from the surrounding cluster. (Note that the argument
would apply equally well to the situation where the entire stellar
content of the system resided in the core, so although we retain
the above nomenclature of `core' and `cluster' we can also envisage a situation
where, for example, the core is actually an entire cluster, located
perhaps at the intersection of inflowing gaseous filaments and thus
subject to a supply of accretable material: see e.g. Bonnell et al 2003).

  In what follows, we omit numerical constants of order unity, since
we apppreciate that our estimates are in any case in need of refinement
by further numerical experiments. In particular, in this simplest analysis
we treat the inner (core) system as being homogeneous (i.e. represented
at any time by a single value of stellar mass, velocity dispersion and
density). The utility of this analysis lies not so much in the
numerical values that we derive but in its setting out of the
underlying principles governing the evolution of an accreting cluster 
core. In particular, we draw attention to the scaling relations 
that we derive for how the maximum stellar density (and hence the
incidence of stellar collisions) depends on the parameters of the parent
cluster.
 
\subsection { The maximum density in the core}
 
 We consider the simple situation of a cluster, mass $M_{cl}$ with a
characteristic dynamical timescale of $t_{dyn}$. If the cluster is
largely gaseous initially, then in the absence of support mechanisms
for this gas, the flow rate of gas into the central regions is

\begin{equation}
\dot M \sim {{M_{clus}} \over {t_{dyn}}}
\end{equation}

 We now consider the case where star formation in the central regions
produces a core of mass $M_{core}$, whose mass grows with time as its
members intercept the accretion flow described by (1). For the purpose
of this simple argument, it is assumed that all the inflowing gas
is accreted by stars in the core, an assumption which is unlikely to
be true in practice once stars attain high masses ($\sim 10 M_\odot$),
due to the effects of feedback (e.g. Wolfire \& Casinelli 1987, Yorke \&
Sonnhalter 2002, Edgar
\& Clarke 2004 ). A consideration of the complex effects
of stellar feedback in an accreting stellar core is beyond the scope
of this Letter: see  e.g. Dale et al 2005 for recent simulations
of this situation. Likewise, we here neglect the possibility that some
of the accretion flow into the core produces new stars via gravitational
instability of the inflowing gas: simulations  suggest that cluster
cores grow in mass by a combination of the infall of small stellar systems
and
the subsequent growth of these stars by further gas accretion: see
Bonnell, Bate and Vine (2003) for a detailed analysis of this issue.

 The timescale on which the core mass grows is thus $t_{\dot M}$:

\begin{equation}
t_{\dot M} \sim {{M_{core}} \over {\dot M}} \sim {{M_{core}} \over{M_{clus}}} t_{dyn}
\end{equation}

Since this is obviously much less than the dynamical timescale of the parent
cluster ($t_{dyn}$), we see that to first order the terms $M_{clus}$ and
$t_{dyn}$ remain roughly constant during core growth and hence $t_{\dot M}$
increases linearly with $M_{core}$.{\bf (We will go on to show below that the density
of the core rises very steeply as the core grows, so that in fact
$ M_{core} << M_{clus}$ throughout the evolution we consider here.
Thus the assumption that $ t_{\dot M} << t_
{dyn}$ also remains valid).}  On the other hand, the {\it core's}
dynamical time ($t_{dyn_{core}}$) decreases with time as the core density
increases and hence the ratio $t_{\dot M}/t_{dyn_{core}}$ increases
monotonically with time. Therefore, the core will at some point enter the
regime where $t_{\dot M}/t_{dyn_{core}}$ becomes larger than unity, and,
as this condition becomes more amply fulfilled with time thereafter, the
core growth becomes, to a better and better approximation, adiabatic.
In this case, the addition of mass on a timescale much longer than the
local dynamical time, results in growth which preserves the adiabatic
invariant $ M_{core} v_{core} R_{core}$,  where  $v_{core}$ and $R_{core}$ are
respectively the velocity dispersion and characteristic radius of the
cluster core. Note that in this regime, individual stellar orbits preserve
their angular momenta. Evidently, our expressions are only valid in the limit
that the angular momentum of the inflowing gas can be neglected, i.e.
to the case that the specific angular momentum of the gas is $<< R_{core} v_{core}$
 ({\bf see  Bonnell \& Bate 2005 for a discussion of the accretion of angular
momentum in the case of a turbulent medium). We here restrict ourselves to
the case of inflow of gas with zero angular momentum in order to make a comparison
with the (non-rotating) simulations of Bonnell \& Bate 2002: evidently
further simulations will be required to test the validity of our analysis
in the case of more complex velocity fields for the gas.} 
If, in addition, the core remains in a state of approximate virial equilibrium
(which is again likely in the adiabatic growth regime), then $v_{core}^2  
\propto
 M_{core}/R_{core}$. Thus, the core mass and radius ($M_{core}$ and
$R_{core}$) are related via the adiabatic relationship:

\begin{equation}
R_{core} \propto M_{core}^{-3}
\end{equation}

  According to this equation, continued accretion drives the core
ineluctably towards higher densities (note that this adiabatic scaling
implies that the mass density in the core scales as $M_{core}^{10}$!).
This runaway increase in core density would then apparently only be halted
by the exhaustion of the accretion flow onto the core (either because
the whole cluster gas mass had accreted onto the core, or because the
supply into the core was limited either by the effects of feedback or by
consumption by star formation in the rest of the cluster).

  However, this inexorable core growth requires that the core shrinkage
timescale ($t_{\dot M}$) is less than the timescale on which the
core is puffed up 
by discrete interactions  
within the core (either   the creation of dynamical binaries and
the associated transfer of kinetic energy into the motions of third bodies
or the aggregate effect of two body encounters).
{\bf The timescale for these effects depends on the number of stars in
the core:
in the case of populous clusters ($N_{core} > 100$), this timescale
is the conventional two body relaxation timescale (see Binney \& Tremaine
1987, eqn. (4-9)), while in the case of low $N_{core}$, this timescale
has 
been quantified by the simulations of
Bonnell \& Clarke 1999 as  $ \sim N t_{dyn_{core}}$ (see also the discussion
in Heggie \& Hut 2003, p. 257).}
\footnote{Note that we are assuming here that this
puffing up is not offset by the outward transference of energy into orbital
motions of stars in the surrounding cluster, i.e. we assume that the
Nbody evolution of the core is dynamically decoupled from that of the
surrounding cluster. Given the runaway growth in core density, it is
probably reasonable to assume such dynamical decoupling. Were this
not the case, the core could be driven to higher densities than we
estimate here}
{\bf We therefore write:  

\begin{equation}
t_{puff} \sim f N_{core} t_{dyn_{core}}
\end{equation}

 where $f =1$ for $N_{core} < 100$ and $f -> 0.1/{\rm ln} N_{core}$
as $N_{core} -> \infty$.}   
 We thus propose that the core can continue to shrink in response to
adiabatic accretion provided that

\begin{equation}
{{M_{core}}\over{M_{clus}}} t_{dyn} < {\bf f} N_{core} t_{dyn_{core}}
\end{equation}

or
\begin{equation}
t_{dyn_{core}}  > {{\bar m} \over { {\bf f} M_{clus}}} t_{dyn}
\end{equation}

where $\bar m_{core}$ is the mean stellar mass in the core.
Since the dynamical timescale is simply proportional to the inverse
square root of the mean enclosed density, it then follows that
the core density and mean cluster density ($\bar \rho $)
are then related via the
inequality

\begin{equation}
\rho_{core} < \biggl({ {\bf f} {M_{clus}}\over{\bar m_{core}}}\biggr)^2 \bar \rho
\end{equation}

%If a fraction $\epsilon$ of the cluster gas eventually goes on to form
%stars with mean mass $\bar m_{clus}$, then the total number of stars
%in the cluster is given by:

%\begin{equation}
%N_* = {{\epsilon M_{clus}} \over {\bar m_{clus}}}
%\end{equation}

%and equation (?) can be written in terms of a maximum core density:

%\begin{equation}
%\rho_{max} = f N_*^2 \bar \rho
%\end{equation}

% where $f=(\bar m_{clus}/\epsilon  \bar m_{core})^2$. We expect for the
%case that the cluster remains bound that $\epsilon > 0.5$, while the
%fact that any observed mass segregation in young clusters is not
%extreme suggests that $\bar m_{core}$ exceeds $\bar m_{clus}$ by a relatively
%modest factor. 
%Thus, although $f$ is not tightly constrained, it is likely to be of order
%unity.  This would mean that the main  dependence of $\rho_{max}$ on
%$\bar \rho$ is via the (square of the) cluster membership number $N_*$.

 We can apply this formula to the simulations of Bonnell \& Bate  2002, for
which $M_{clus} \sim 10^3 M_\odot$ 
(since the simulation was designed to mimic
the Orion Nebula Cluster) and where the mean stellar mass in the core
is $\sim 3 M_\odot$. We find, by inspection of Figure 2 of Bonnell \&
Bate,
that
the core density indeed increases by about five  orders of magnitude
compared with the mean cluster density. Bonnell \& Bate
found that further
shrinkage of the core was at this stage offset by puffing up due to few
body interactions in the dynamically decoupled cluster core.

\section{The mean number of collisions per star.}

  The number of collisions experienced by a star is given by the product
of the collision rate at the highest density achieved by the stellar core
and the lifetime of the core in that highest density state. We have argued above that this latter is
given by the timescale on which the core  puffs up by few-body effects which,
in the case that the velocity dispersion in the core is $v_{core}$,  
is approximately equal to the timescale on which a star experiences a
gravitationally focused  encounter with another
star, i.e. an encounter with a star within impact parameter:

\begin{equation}
r_{enc} = {{G \bar m}\over {v_{core}^2}}. 
\end{equation}

Since stellar collisions result from situations where the closest approach
distance $r_{coll}$ is generally $<< r_{enc}$, and since all encounters within
$r_{enc}$ are, by definition, in the gravitationally focused regime, we
can  write the fraction of gravitationally focused encounters
that lead to collision as the ratio of $r_{coll}$ to $r_{enc}$.
 The mean number of collisions per star is thus given by

\begin{equation}
\bar n_{coll} = {{r_{coll}} \over {r_{enc}}}
\end{equation}

\noindent where we here define $r_{coll}$ as the separation of closest approach which would
lead, directly or indirectly, to a stellar collision. The indirect
route to stellar collisions involves the interaction between a passing
star and a hard binary such that, during the interaction, two of
the stars pass within a stellar radius of each other.
From the three-body computations of Davies, Benz \& Hills (1994), one
may deduce that in the case of gravitationally focused encounters for which the
intruder's nominal closest approach to the binary (if the binary were
treated as a point mass) was of order the  binary separation - the
minimum encounter distance is commonly about a tenth of the
binary separation. Hence  we would expect that 
stellar collisions
could result from such encounters in the case of binaries with
separation of about $10$ stellar radii (or about $1$ A.U.).
Thus provided massive stars in the cluster core are frequently located in
binaries of this separation
(consistent with the observed binary properties of 
OB stars in general: see Mason et al 1998, Garcia \& Mermillliod 2001) then 
$r_{coll}$ may exceed $r_*$ (the stellar radius) by an order of magnitude.
If we define $v_{coll}$ as the escape velocity at separation $r_{coll}$ (rather
than $r_*$), 
we can then rewrite (9) as:

\begin{equation}
\bar n_{coll} = \biggl({{v_{core}} \over {v_{coll}}}\biggr)^2
\end{equation}

 In order to be useful, we need to relate $v_{core}$, the velocity dispersion
in the core at its maximum density, to the velocity dispersion of the
parent cluster, $\tilde v$. A condition of approximate virial equilibrium
implies that velocity dispersion scales with system mass and density as
$\propto M^{1/3}\rho^{1/6}$, so that, at maximum density, equation (7)
implies that 

\begin{equation}
{{v_{core}} \over {\tilde v}} \sim  \bigl( {\bf f} N_{core} \bigr)^{1/3}
\end{equation}

  Thus the mean number of collisions per star in the core is given by:

\begin{equation}
\bar n_{coll} \sim \bigl({\bf f} N_{core}\bigr)^{2/3} \biggl({{\tilde v} \over {v_{coll}}}\biggr)^2
\end{equation}
 Likewise, the total number of collisions expected in the core is:

\begin{equation}
\bar N_{coll} \sim \bf f^{2/3} N_{core}^{5/3} \biggl({{\tilde v} \over {v_{coll}}}\biggr)^2
\end{equation}

  These equations allow one to assess which clusters  are suitable
for producing  accretion induced stellar collisions. If we invoke a high
binary fraction, so that $r_{coll} \sim 10 r_*$, then $(\tilde v /v_{coll})^2 $
is typically $\sim 10^{-3}-10^{-2}$ for clusters  with $\tilde v$ in the
range $5-20$ km s$^{-1}$. This places a lower limit of $N_{core} \sim 20-100$
before one expects collisions to occur, and a lower limit of
$N_{core} \sim 10^3-3 \times 10^4$ in order for the majority of stars
in the core to be involved in collisions.

   We thus see that - given that $v_{coll}$ is just set by the  stellar
mass-radius
relationship and the binary fraction among massive stars - the main 
{\it environmental} determinants
of whether stellar collisions are expected are the velocity dispersion,
$\tilde v$, and the membership number, $N_{core}$ of the adiabatically
contracting core. This raises the obvious question of how the value of
$N_{core}$ relates to the parameters of the larger system. Clearly,
once a sub-system starts to contract adiabatically, its subsequent
evolution becomes increasingly adiabatic. Simple application of the
adiabatic condition, suggests that a sub-system  will
embark on the path of adiabatic contraction if its internal velocity
dispersion exceeds that  of the parent cluster (gas reservoir),
although this hypothesis has not been subject to numerical investigation.
In this case, the existence and scale of any adiabatic core 
would depend on the details of the cluster's radial density profile.
We note that our simple one zone model for the evolution of the 
adiabatic core does not describe the possibility that the number of
stars involved in adiabatic evolution may increase during the collapse. Evidently,
this is an issue that requires numerical investigation.
It is thus difficult, pending further simulations, to
decide how $N_{core}$ relates to the mass of the parent cluster, apart from
the obvious expectation that the total number of stars in the cluster
$N_*$ should comfortably exceed $N_{core}$. 

\section{Conclusions}

  We have set up a simple theoretical model which demonstrates why 
current simulations of accretion of gas onto the cores of
stellar clusters do not drive stellar densities to the point where 
stellar collisions occcur. This is because the conditions modeled -
motivated by the properties of the Orion Nebula Cluster - involve free
fall velocities of a few km s $^{-1}$ and - with a total cluster mass
of around $10^3 M_\odot$ - the development of an adiabatic core containing
only a few tens of stars. According to equation (13), the expected number of
collisions in such a system is $\sim 0.01$. 
With such a low free fall velocity the core's shrinkage timescale
is relatively long, while, with a core containing such a low number of
stars, the core's  dissolution timescale due to Nbody effects is 
relatively short. Consequently, the system cannot shrink to the densities
required for stellar collisions.

  We find however that the number of collisions is strongly dependent on
both the free fall velocity of the parent cluster ($\tilde v$) and the number of stars in the
adiabatic core ($N_{core}$), scaling as $N_{core}^{5/3} \tilde v^2$.
Hence, even without increasing the number of stars in the core, an
increase in $\tilde v$ to $\sim 20$ km s$^{-1}$ increases the expected number
of collisions to of order unity. Since it is also likely that $N_{core}$
would be larger for a more massive over-all system, it would appear that
stellar collisions are a likely outcome in  more populous, more tightly
bound clusters such as globular clusters or super star clusters.

  We note that the scenario outlined in this paper bears comparison
with another scenario that leads to stellar collisions in an ultra-dense
cluster core, namely the action of the Spitzer (mass segregation ) instability,
(see  Portegies Zwart \& McMillan 2002, G\"urkan et al 2004, Portegies Zwart et al 2004 
and Freitag et al 2006a),b)).
In both cases, the onset of stellar collisions raises the possibility of
producing supermassive stars and stellar remnants, the latter being discussed
as possible seed black holes for eventual merger into supermassive
black holes in galactic nuclei.

 In the case of the Spitzer instability, 
high densities are achieved by two body relaxation
in the presence of a stellar mass spectrum and the limiting factor is 
whether the timescale for this process is sufficiently short so as to
allow collisions to occur over the (few Myr) timescale on which the
most massive (largest cross-section) stars expire as a supernova.
In the present case, the timescale for core shrinkage is of order
the cluster dynamical timescale ($\sim 10^5$ years) and therefore collapse
is obviously fast enough compared with stellar evolutionary timescales.
Here, instead, the issue hinges on the maximum density that is achievable
and whether adiabatic shrinkage is offset by either two body effects 
(as discussed here) or by the quenching of accretion into the cluster core.
Although detailed consideration
of this latter effect is beyond the scope of this paper, we here  note that the
efficacy of feedback by ionising radiation is highly (and negatively) 
dependent on gas density (Dale et al 2005). 
This factor thus reinforces the fact that effective shrinkage of
the cluster core is favoured in the densest clusters.

\section{Acknowledgments}
We are grateful to Marc Freitag for a critical reading
of the manuscript and for useful discussions. {\bf We thank the 
referee for helping us to clarify some of our arguments}.

%\section{References}

%\begin{references}

\end{document}